# The collision between two hydrogen atoms

Hasi Ray[1,2,3]

[1] Science Department, National Institute of TTT and Research, Kolkata 700106, India
[2] Department of Physics, New Alipore College, Kolkata 700053, India
[3] Indian Association for the Cultivation of Science, Jadavpur, Kolkata 700032, India

Email: hasi_ray@yahoo.com

**Abstract**: The electron-electron correlation term in two-atomic collision is the most important, most difficult term to obtain the effective interatomic potential. Generally the H and H collision is a four center problem. It is extremely difficult to compute the electron-electron correlation term to include the effect of exchange or antisymmetry between two system electrons exactly. All the two-atomic collision related theoretical data differ from each other due to its difference in approximating the electron-electron correlation term. I invent a trick to evaluate the term exactly. Earlier the positronium (Ps) and H system was easily approximated as a three center problem due to the light mass of Ps. My new code for H-H collision using the ab-initio and exact static-exchange model (SEM) can reproduce exactly the same data of Ps and H system just by using the appropriate atomic parameters. The success of the present trick makes the foundation of a big monument in cold and low energy atomic collision physics. The Feshbach resonances are observed in H(1s)-H(1s) elastic scattering in both the singlet and triplet channels. One can use the trick to solve any four center problem including excitation and ionization processes. Again one can use the idea in larger systems too.

PACS 02.90.+p – Other topics in Mathematical methods

PACS 03.65.Nk – Scattering theory

PACS 34.20.Cf – Interatomic potential and forces

PACS 34.50.Cx – Elastic; ultracold collisions

The importance of two-atomic collision is well known since the beginning [1] and today its importance is at the top to know the basic sciences and its application in technology [2]. The appropriate qubit generation for quantum computation is a challenging job today [3]. Recently the scientific community realizes the need of quantum control of isolated atomic system [4] with nanoscale localization instead of the collective motion of atoms as new possibilities for quantum control [5]. The controlling of the s-wave scattering length ($a$) in ultracold collision is an important subject of study to tune the Feshbach resonances [6]. The simplest and ideal two-atomic collision system is the H-H system. Here both the nuclei are heavy and it is strictly a four center problem. It is extremely difficult to compute the electron-electron correlation term in H-H collision to include the effect of exchange or antisymmetry between two system electrons exactly. It is relatively less difficult in positronium (Ps) and H system [7]; due to the very light mass of Ps it is a good approximation to consider the H-nucleus as the origin of center of mass system in Ps-H system and so it was treated [7] as a three center problem. But in a system of two heavy atoms like H-H, Li-H, Li-Li etc. or two very light exotic atoms like Ps-Ps, Muonium (Mm) –Mm, it is quite difficult. I invent a trick to evaluate the electron-electron Coulomb-correlation term including the exchange between them exactly. The new code for two-atomic collision system using the ab-initio and exact static-exchange model (SEM) [7] can reproduce exactly the same data of Ps-H system just by manipulating the parameters of the atomic wavefunctions. The success of the present idea makes the foundation of a big monument in the cold and low energy atomic collision physics. In addition one can use the trick to solve any four center problem even to calculate excitation and ionization amplitudes and can extend the trick in larger systems too.

To obtain the two-atomic collision amplitude, we use a coupled-channel method [8] with an eigen-state expansion in the basis states to solve the Schrodinger equation. The unknown coefficients of the eigen-state expansion carry the informations of the scattering process. It is a function of the incident energy and the scattering angle and known as the scattering amplitude. We use the SEM [7,9,10] theory when both the atoms are at ground states to study the elastic collision:

$$A(n,l,m) + B(n',l',m') \Rightarrow A(n,l,m) + B(n',l',m') \quad (2.1)$$

if $A$ and $B$ are two atoms and $|\vec{k}_i| = |\vec{k}_f|$ if $\vec{k}_i$ and $\vec{k}_f$ represent the initial and final momenta of the projectile. Only the direction of the final momentum $\vec{k}_f$ changes due to elastic scattering. The center of mass (CM) frame is used. The initial and final state wavefunctions are defined as

$$\psi_i = e^{i\vec{k}_i \cdot \vec{R}'} \phi^A_{1s}(\vec{r}_1) \phi^B_{1s}(\vec{r}_2) \quad (2.2)$$

$$\psi_f = (1 \pm P_{12}) e^{i\vec{k}_f \cdot \vec{R}_f} \phi^A_{1s}(\vec{r}_1) \phi^B_{1s}(\vec{r}_2) \quad (2.3)$$

Here $\phi^A_{1s}(\vec{r}_1)$ and $\phi^B_{1s}(\vec{r}_2)$ are ground state wave functions of the two atoms and $P_{12}$ is the exchange (or antisymmetry) operator. The electron spins are taken into consideration explicitly. The plus (+) sign is used in the space part to represent the singlet (S=0) channel and the minus (-) sign indicates the triplet (S=1) spin-configurations.

Projecting different states on the Schrodinger equation just like the Hartree-Fock variational method one can get the integro-differential equations that can be solved by the method of iteration. We use the Lippman-Schwinger type integral equation in the momentum space formalism [8] rather than using the coordinate space adapted by Fraser in Ref [11]. The formally exact Lippman-Schwinger type coupled integral equation for the scattering amplitude in momentum space is given by [8]:

$$f^{\pm}_{n'1s,n1s}(\vec{k}_f,\vec{k}_i) = B^{\pm}_{n'1s,n1s}(\vec{k}_f,\vec{k}_i) - \frac{1}{2\pi^2} \sum_{n''} \int d\vec{k}'' \frac{B^{\pm}_{n'1s,n''1s}(\vec{k}_f,\vec{k}'') f^{\pm}_{n''1s,n1s}(\vec{k}'',\vec{k}_i)}{\vec{k}^2_{n''1s} - \vec{k}''^2 + i\varepsilon} \quad (2.4)$$

Here $B^{\pm}$ are the well known Born-Oppenheimer (BO) scattering amplitude [9,10] in the singlet (+) and triplet (-) states respectively. In a similar fashion, $f^{\pm}$ indicate the unknown scattering amplitudes for the singlet and triplet states of the two system electrons. Generally the partial wave analysis is used to reduce the three-dimensional integral equation into the one-dimensional form. Here the BO amplitude ($B^{\pm}$) acts as the input to get the SEM amplitude following the Eqn. (2.4) and is defined as

$$B^{\pm}_{n'1s,n1s}(\vec{k}_f,\vec{k}_i) = -\frac{\mu}{2\pi}\int d\vec{R}d\vec{r}_1 d\vec{r}_2 \psi_f^*(\vec{R},\vec{r}_1,\vec{r}_2) V(\vec{R},\vec{r}_1,\vec{r}_2) \psi_i(\vec{R},\vec{r}_1,\vec{r}_2) \quad (2.5)$$

When $\mu$ is the reduced mass of the system and $V(\vec{R},\vec{r}_1,\vec{r}_2)$ is the Coulomb interaction: $V_{Direct}$ is for the direct channel and $V_{Exchange}$ is for the exchange or rearrangement channel.

$$V_{Direct}(\vec{R},\vec{r}_1,\vec{r}_2) = \frac{Z_A Z_B}{R} - \frac{Z_A}{|\vec{R}-\vec{r}_2|} - \frac{Z_B}{|\vec{R}+\vec{r}_1|} + \frac{1}{|\vec{R}+\vec{r}_1-\vec{r}_2|} \quad (2.6a)$$

$$V_{Exchange}(\vec{R},\vec{r}_1,\vec{r}_2) = \frac{Z_A Z_B}{R} - \frac{Z_A}{|\vec{r}_1|} - \frac{Z_B}{|\vec{r}_2|} + \frac{1}{|\vec{R}+\vec{r}_1-\vec{r}_2|} \quad (2.6b)$$

The atomic unit (a.u.) is used throughout. The four Coulomb interaction terms: the first one is the nucleus-nucleus (NN) interaction, the fourth one is the electron-electron ($e_1 e_2$) interaction, the second one is the interaction between nucleus A and electron 2 (Ae), and the third one is the interaction between nucleus B and electron 1 (Be). All the four interactions are taken into consideration with exact exchange to calculate the Born-Oppenheimer scattering amplitude $B^{\pm}$.

The explicit form of the first term in the direct ($F_B^{NN}$) and rearrangement ($F_O^{NN}$) channels are:

$$F_B^{NN} = -\frac{\mu}{2\pi}\int d\vec{R}d\vec{r}_1 d\vec{r}_2 e^{-i\vec{k}_f \cdot \vec{R}} \phi_{1s}^{A*}(\vec{r}_1) \phi_{1s}^{B*}(\vec{r}_2) \frac{Z_A Z_B}{R} e^{i\vec{k}_i \cdot \vec{R}} \phi_{1s}^A(\vec{r}_1) \phi_{1s}^B(\vec{r}_2) \quad (2.7a)$$

$$F_O^{NN} = -\frac{\mu}{2\pi}\int d\vec{R}d\vec{r}_1 d\vec{r}_2 e^{-i\vec{k}_f \cdot \vec{R}_f} \phi_{1s}^{A*}(\vec{R}-\vec{r}_2) \phi_{1s}^{B*}(\vec{R}+\vec{r}_1) \frac{Z_A Z_B}{R} e^{i\vec{k}_i \cdot \vec{R}} \phi_{1s}^A(\vec{r}_1) \phi_{1s}^B(\vec{r}_2) \quad (2.7b)$$

Similarly the fourth electron-electron correlation terms are,

$$F_B^{e_1 e_2} = -\frac{\mu}{2\pi}\int d\vec{R}d\vec{r}_1 d\vec{r}_2 e^{-i\vec{k}_f \cdot \vec{R}} \phi_{1s}^{A*}(\vec{r}_1) \phi_{1s}^{B*}(\vec{r}_2) \frac{1}{|\vec{R}+\vec{r}_1-\vec{r}_2|} e^{i\vec{k}_i \cdot \vec{R}} \phi_{1s}^A(\vec{r}_1) \phi_{1s}^B(\vec{r}_2) \quad (2.8a)$$

$$F_O^{e_1 e_2} = -\frac{\mu}{2\pi}\int d\vec{R}d\vec{r}_1 d\vec{r}_2 e^{-i\vec{k}_f \cdot \vec{R}_f} \phi_{1s}^{A*}(\vec{R}-\vec{r}_2) \phi_{1s}^{B*}(\vec{R}+\vec{r}_1) \frac{1}{|\vec{R}+\vec{r}_1-\vec{r}_2|} e^{i\vec{k}_i \cdot \vec{R}} \phi_{1s}^A(\vec{r}_1) \phi_{1s}^B(\vec{r}_2) \quad (2.8b)$$

The second and third terms in direct and exchange channels are:

$$F_B^{Ae} = -\frac{\mu}{2\pi}\int d\vec{R}d\vec{r}_1 d\vec{r}_2 e^{-i\vec{k}_f \cdot \vec{R}} \phi_{1s}^{A*}(\vec{r}_1) \phi_{1s}^{B*}(\vec{r}_2) \frac{(-Z_A)}{|\vec{R}-\vec{r}_2|} e^{i\vec{k}_i \cdot \vec{R}} \phi_{1s}^A(\vec{r}_1) \phi_{1s}^B(\vec{r}_2) \quad (2.9a)$$

$$F_O^{Ae} = -\frac{\mu}{2\pi}\int d\vec{R}d\vec{r}_1 d\vec{r}_2 e^{-i\vec{k}_f \cdot \vec{R}_f} \phi_{1s}^{A*}(\vec{R}-\vec{r}_2) \phi_{1s}^{B*}(\vec{R}+\vec{r}_1) \frac{(-Z_A)}{|\vec{r}_1|} e^{i\vec{k}_i \cdot \vec{R}} \phi_{1s}^A(\vec{r}_1) \phi_{1s}^B(\vec{r}_2) \quad (2.9b)$$

$$F_B^{Be} = -\frac{\mu}{2\pi}\int d\vec{R}d\vec{r}_1 d\vec{r}_2 e^{-i\vec{k}_f \cdot \vec{R}} \phi_{1s}^{A*}(\vec{r}_1) \phi_{1s}^{B*}(\vec{r}_2) \frac{(-Z_B)}{|\vec{R}+\vec{r}_1|} e^{i\vec{k}_i \cdot \vec{R}} \phi_{1s}^A(\vec{r}_1) \phi_{1s}^B(\vec{r}_2) \quad (2.10a)$$

$$F_O^{Be} = \frac{\mu}{2\pi}\int d\vec{R}d\vec{r}_1 d\vec{r}_2 e^{-i\vec{k}_f \cdot \vec{R}_f}\phi_{1s}^{A*}(\vec{R}-\vec{r}_2)\phi_{1s}^{B*}(\vec{R}+\vec{r}_1)\frac{(-Z_B)}{|\vec{r}_2|}e^{i\vec{k}_i \cdot \vec{R}}\phi_{1s}^{A}(\vec{r}_1)\phi_{1s}^{B}(\vec{r}_2) \quad (2.10b)$$

with
$$\vec{R}' = \vec{R} + \frac{m_e}{m_A + m_e}\vec{r}_1 - \frac{m_e}{m_B + m_e}\vec{r}_2 \quad (2.11a)$$

$$\vec{R}_f = \vec{R} + \frac{m_e}{m_A + m_e}(\vec{r}_2 - \vec{R}) - \frac{m_e}{m_B + m_e}(\vec{r}_1 + \vec{R}) \quad (2.11b)$$

Here $\vec{R}$ is the inter-nucleus separation, $\vec{r}_1$ and $\vec{r}_2$ are the position vectors of the two system electrons with respect to their corresponding atomic nuclei. The notations $m_A$, $m_B$, $m_e$ represent the masses of the nucleus A, nucleus B and the mass of electron. The most difficult term is the integral in Eqn. (2.8b) i.e. electron-electron correlation with exact exchange; it contains three completely different coupled terms and two uncoupled terms in a nine-dimensional integral. I invent a trick to transform this integral into an easily tractable two-dimensional form using no approximation, with a simple substitution and accordingly changing the nine-dimensional space of integration.

To determine the s-wave elastic scattering lengths ($a^+$ and $a^-$), the effective range theory that expresses s-wave elastic phase shift ($\delta_0$) as a function of scattering-length ($a$) and projectile energy ($\sim k^2$) so that

$$k \cot \delta_0 = -\frac{1}{a} + \frac{1}{2}r_0 k^2 \quad (2.12)$$

is used; when $k$ is the magnitude of the incident momentum in atomic unit; $r_0$ is the range of the potential. The scattering length, $a \langle 0$ indicates the possibility of no binding in the system. Only the positive scattering-length i.e. $a \rangle 0$ indicates the possibility of binding, the Feshbach resonances and the BEC. A rapid change in phase-shift by $\pi$ radian is an indication of the presence of a Feshbach resonance [10,12] in the system; the presence of such a resonance indicates the possibility of a binding.

A highly efficient computer code is developed using the FORTRAN programming language and numerical analysis. The partial wave analysis is used in such a way that L=0 indicates s-wave, L=1 indicates p-wave and so on. The code calculates the elastic phase-shifts and corresponding cross sections for both the singlet (S=0) and the triplet (S=1) spin-states of the the two system electrons. Two sets of coupled one-dimensional integral equation one corresponding to $f_L^+$ and the other $f_L^-$, are solved numerically by the matrix inversion method for each partial wave (L). The BO amplitudes are evaluated in two ways as a check of the method. In addition, the values of the integrated BO cross sections evaluated by partial wave and non-partial wave method are found to be identical. The convergence of the required scattering amplitudes are tested by increasing the number of Gaussian quadratures in solving the integral equations. The pole term evaluated using the formulation with delta function and principal value parts is as follows:

$$\frac{1}{\vec{k}_n^2 - \vec{k}''^2 + i\varepsilon} = -i\pi\delta(\vec{k}_n^2 - \vec{k}''^2) + \frac{P}{\vec{k}_n^2 - \vec{k}''^2} \quad (2.13)$$

The principal value integral from zero to infinity has been replaced by

$$\int_0^\infty dk'' = \int_0^{2k_n} dk'' + \int_{2k_n}^\infty dk'' \qquad (2.14)$$

Even number of Gaussian points in the interval $0-2k_n$ are used to avoid the singularity problem at $k'' = k_n$. To include the contributions of the higher partial waves as accurately as possible in the integrated cross sections, an augmented Born approximation [7,9,10] is used in the new code.

In the present new code there are four parameters of the atomic wavefunctions, two from each atom, used as inputs. The two parameters of an atom are the Coulomb-screening parameter ($\lambda$) and the radius of the electronic orbit. To verify the correctness of the code, the appropriate atomic parameters for H-Ps system is used in the present code and it has reproduced exactly the same data [7]. I use the code to study H(1s)-H(1s) elastic scattering in search of Feshbach resonances. A very large number of energy mesh-points are used in the energy interval $E = 1.4 \times 10^{-4} eV$ to $E = 1.2 \times 10^{-2} eV$ to obtain s-wave elastic phase-shifts and the corresponding cross sections. In **Figures 1** and **2**, the s-wave (L-0) elastic phase-shift and partial cross section data of H-H scattering using the present SEM code for both the singlet (S=0) and triplet (S=1) spin configurations of the system electrons are plotted against the incident momenta, $k = 0.1$ to $k = 0.9$ a.u. The incident energy is related to $k$ by the relation $E(eV) = (27.21 k^2 / 2\mu)$. Many Feshbach resonances are observed in the singlet (+) channel and only one Feshbach resonance is found in the triplet (-) channel at the incident energy $E \sim 4.5 \times 10^{-3} eV$. Here both the partial phase-shift and the partial cross section are computed directly from the real and imaginary parts of the s-wave scattering amplitudes for the singlet ($f_{L=0}^+$) and the triplet ($f_{L=0}^-$) channels. The tabular data for s-, p- and d-wave phase-shifts and corresponding partial cross sections are presented in **Tables 1 and 2**. It is very difficult to get singlet scattering length ($a^+$) due to presence of many resonances in the singlet channel.

To get the triplet scattering length ($a^-$), the $k \cot \delta_0$ is plotted against $k^2$ in **Figure 3** following effective-range theory. To ascertain the correctness of the present code, similar data of H-Ps and Ps-Ps systems obtained using the present four-center code are compared in the same **figure 3** with the data of H-H collision system. In **Table 3**, the computed scattering lengths are presented and compared with available theoretical data. The triplet scattering length of H-Ps system obtained by using the present code is in good agreement with available data; the scattering lengths of Ps-Ps and H-H system are also close to the available data. Again the integrated elastic cross sections of the H-H scattering using the augmented-Born approximation in SEM are presented and compared with the integrated cross sections of H-Ps system in **Figure 4**.

In conclusion, an exact and ab-initio static-exchange model and computer-code is developed to study two-atomic collision. A simple and efficient trick in invented to evaluate the electron-electron Coulomb-correlation term with exchange exactly in a four-center collision system. The H(1s)-H(1s) elastic scattering is studied using the code. Many Feshbach resonances are observed in the singlet channel and only one Feshbach resonance is observed in the triplet channel at the energy $E \sim 4.5 \times 10^{-3} eV$ and the momentum $k \sim 0.55$ a.u. One can apply the trick to study other systems and not only the elastic scattering but other processes like excitation and ionization etc. It is also possible to use the idea in larger systems. The present code can be useful to study the collision between any two hydrogen-like atoms. Again it is possible to use the code to study alkali atomic collision just by a little modification. My data of H-H, Ps-H and Ps-Ps scattering reported here using static-exchange model are almost absolute except a negligible numerical error due to computation. It is useful to extend the eigen-state expansion basis and a coupled-channel methodology to improve the present code for better accuracy of the data.

The author would be grateful to acknowledge the financial support through Grant No. SR/WOSA/PS-13/2009 of Department of Science & Technology (DST), India.

**References:**

1. H. S. W. Massey and C. B. O. Mohr, Proc. Phys. Soc. **67**, 695 (1954); P. G. Burke, H. M. Schey and K. Smith, Phys. Rev. **129**, 1258 (1963).

2. J. Weiner, V. S. Bagnato, S. Zilio and P. S. Julienne, Rev. Mod. Phys. **71**, 1-85 (1999); W. Ketterle, Rev. Mod. Phys. **74**, 1131-1151 (2002).

3. L. DiCarlo, J. M. Chow, J. M. Gambetta, Lev S. Bishop, B. R. Johnson, D. I. Schuster, J. Majer, A. Blais, L. Frunzio, S. M. Girvin, R. J. Schoelkopf, Nature **460,** 7252 (2009).

4. M. Gullans, T. G. Tiecke, D. E. Chang, J. Fiest, J. D. Thompson, J. I. Cirac, P. Zoller and M. D. Lukin, Phys. Rev. Lett. **109**, 235309 (2012).

5. J. Mizrahi, C. Senko, B. Neyenhuis, K. G. Johnson, W. C. Campbell, C. W. S. Conover and C. Monroe, Phys. Rev. Lett. **110**, 203001 (2013).

6. Ludovic Pricoupenko, Phys. Rev. Lett. **110**, 180402 (2013).

7. H. Ray and A. S. Ghosh, J. Phys. B **29**, 5505 (1996); ibid **30**, 3745 (1997); J. Phys. B **31**, 4427 (1998).

8. A. S. Ghosh, N. C. Sil and P. Mondal, Phys. Rep. **87**, 313-406 (1982).

9. H. Ray, J. Phys. B **32**, 5681 (1999); ibid **33**, 4285 (2000); ibid **35**, 2625 (2002).

10. H. Ray, Phys. Rev. A **73**, 064501 (2006); H. Ray, GSFC NASA Conference Proceeding on Atomic and Molecular Physics edited by A. K. Bhatia, p. 121, published on January 2007 (NASA/CP-2006-214146).

11. P. A. Fraser, Proc. Roy. Soc. B **78**, 329 (1961); J. Phys. B. **1**, 1006 (1968).

12. R. J. Drachman and S. K. Houston, Phys. Rev. A **12,** 885 (1975); ibid **14,** 894 (1976); B. A. . Page, J. Phys. B **9,** 1111 (1976); C. P. Campbell, M. T. McAlinden, F. G. R. S. McDonald and H. R. J. Walters, Phys. Rev. Lett. **80**, 5097 (1998).

13. A. Sen, S. Chakraborty and A. S. Ghosh, Europhys. Lett. **76**, 582 (2006).

14. N. Koyama and J. C. Baird, J. Phys. Soc. Japan **55**, 801 (1986).

15. M. J. Jamieson, A. Dalgarno and J. N. Yukich, Phys. Rev. A **46**, 6956 (1992).

16. C. P. Campbell, M. T. McAlinden, F. G. R. S. MacDonald and H. R. J. Walters, Phys. Rev. Lett. 80, 5097 (1998).

17. P. K. Sinha, A. Basu and A. S. Ghosh, J. Phys. B **33**, 2579 (2000).

18. J. Shumway and D. M. Caperley, Phys. Rev. B **63**, 165209 (2001).

19. K. Oda, T. Miyakawa, H. Yabu and T. Suzuki, J. Phys. Soc. (Japan) **70**, 1549 (2001).

20. I. A. Ivanov, J. Mitroy and K. Varga, Phys. Rev. Lett. **87**, 063201 (2001).

21. S. K. Adhikari, Phys. Lett. A **294**, 308 (2002).

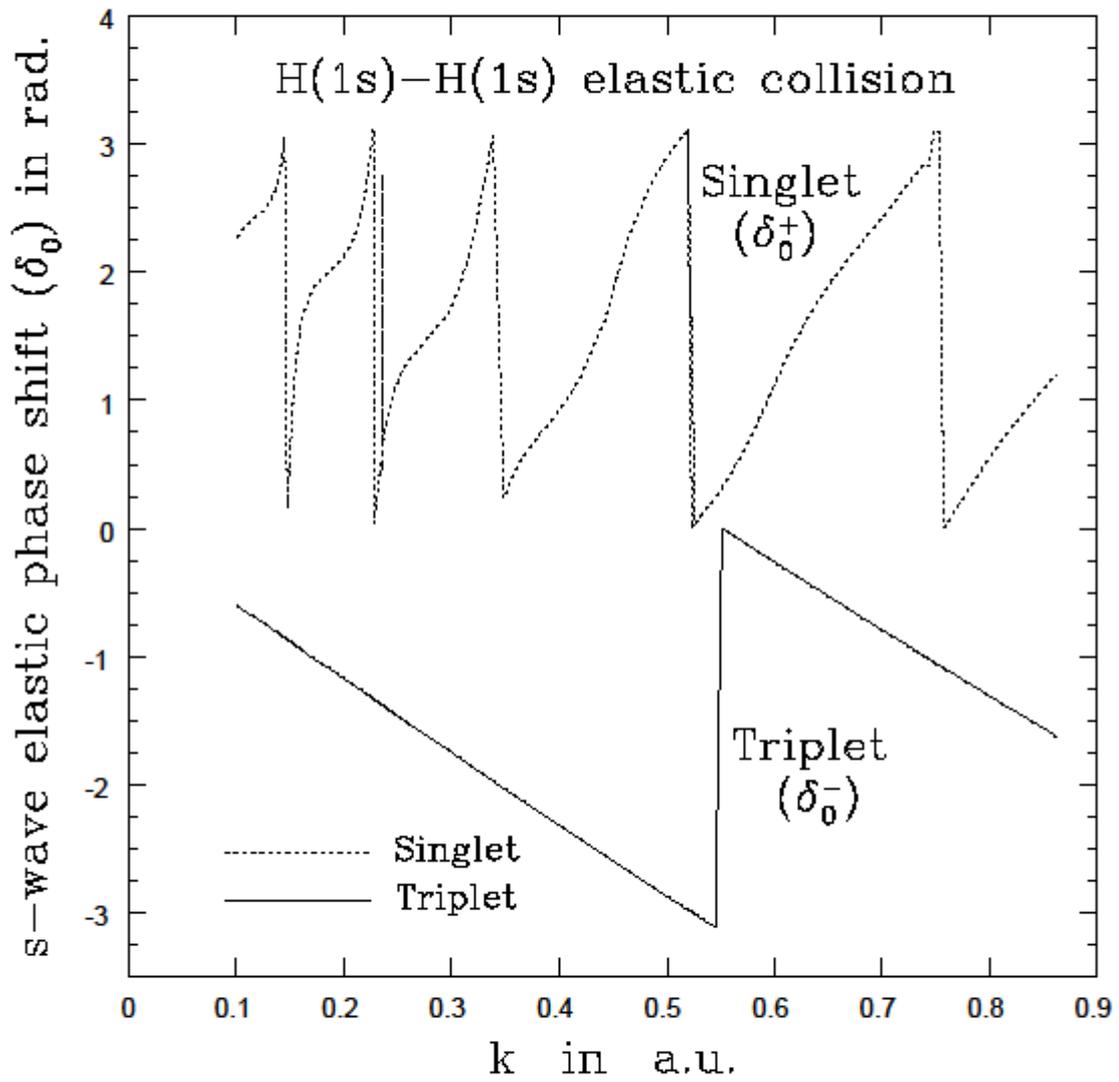

**Figure 1.** The s-wave elastic phase shifts in radian for both singlet (+) singlet and triplet (-) channels in H-H scattering is plotted against the values of incident momentum k in a.u..

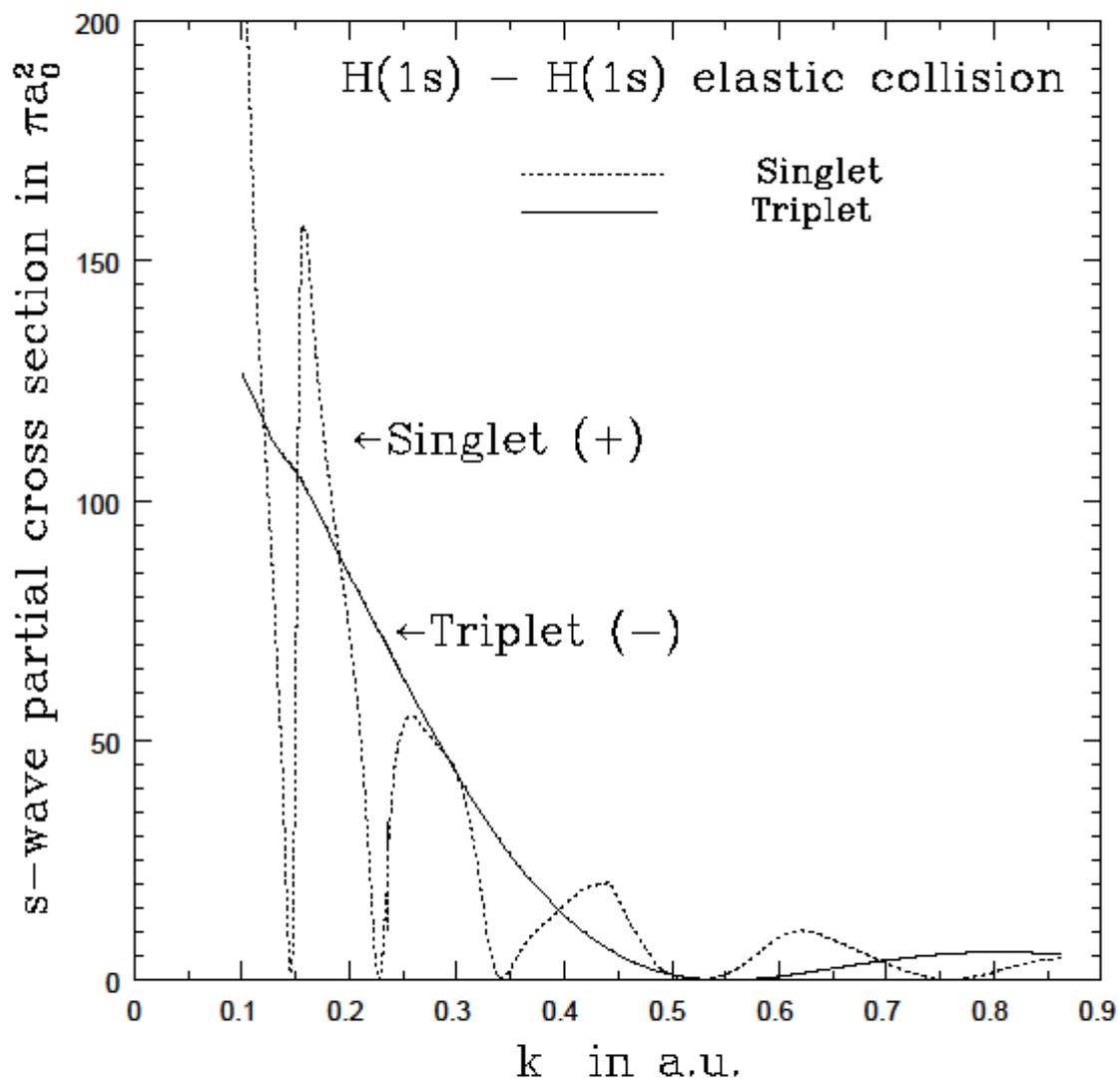

**Figure 2.** The s-wave elastic cross sections in $\pi a_0^2$ for both singlet (+) singlet and triplet (-) channels in H-H scattering is plotted against the values of incident momentum k in a.u.

**Table 1.** The s-, p- and d- wave phase shifts in radian for both singlet (+) and triplet (-) channels in H(1s) and H(1s) elastic scattering.

| k (a.u.) | Singlet phase-shift in radian H and H system | | | Triplet phase-shift in radian H and H system | | |
|---|---|---|---|---|---|---|
| | s-wave | p-wave | d-wave | s-wave | p-wave | d-wave |
| 0.1 | 2.265 | 3.119 | 3.136 | -0.598 | -0.056 | -0.002 |
| 0.2 | 2.122 | 2.441 | 0.001 | -1.167 | -0.312 | -0.036 |
| 0.3 | 1.732 | 1.988 | 0.033 | -1.743 | -0.703 | -0.174 |
| 0.4 | 0.916 | 0.876 | 2.870 | -2.310 | -1.158 | -0.430 |
| 0.5 | 2.896 | 2.325 | 2.118 | -2.866 | -1.640 | -0.774 |
| 0.6 | 1.118 | 0.922 | 0.485 | -0.265 | -2.134 | -1.167 |
| 0.7 | 2.421 | 2.326 | 1.768 | -0.794 | -2.628 | -1.587 |
| 0.8 | 0.551 | 0.303 | 0.089 | -1.311 | -3.117 | -2.022 |
| 0.9 | 1.567 | 1.467 | 1.115 | -1.814 | -0.459 | -2.461 |

**Table 2.** The s-, p- and d- wave cross sections in $\pi a_0^2$ for both singlet (+) and triplet (-) channels in H(1s) and H(1s) elastic scattering.

| k (a.u.) | Singlet (+) cross-section in $\pi a_0^2$ H - H system | | | Triplet (-) cross-section in $\pi a_0^2$ H - H system | | |
|---|---|---|---|---|---|---|
| | s-wave | p-wave | d-wave | s-wave | p-wave | d-wave |
| 0.1 | 236.131 | 0.619 | 0.050 | 126.716 | 3.759 | 0.005 |
| 0.2 | 72.529 | 124.678 | 0.001 | 84.581 | 28.319 | 0.649 |
| 0.3 | 43.301 | 111.466 | 0.248 | 43.133 | 55.662 | 6.632 |
| 0.4 | 15.727 | 44.249 | 8.973 | 13.650 | 62.931 | 21.768 |
| 0.5 | 0.942 | 25.509 | 58.335 | 1.187 | 47.767 | 39.068 |
| 0.6 | 8.987 | 21.15 | 12.091 | 0.765 | 23.840 | 46.988 |
| 0.7 | 3.550 | 12.973 | 39.252 | 4.152 | 5.915 | 40.805 |
| 0.8 | 1.712 | 1.669 | 0.247 | 5.837 | 0.011 | 25.311 |
| 0.9 | 4.938 | 14.656 | 19.907 | 4.650 | 2.904 | 9.766 |

**Table 3.** The triplet (-) scattering lengths in a.u. for elastic scattering of different two-atomic systems.

| Different systems | The triplet (-) scattering length ($a^-$) in a.u. | |
|---|---|---|
| | Present results | Data of others |
| H(1s)-H(1s) | 5.71 | 5.90 [13], 1.3[14], 1.91 [15] |
| H(1s)-Ps(1s) | 2.47 | 2.49 [16], 2.45 [17] |
| Ps(1s)-Ps(1s) | 3.26 | 3.0 [18], 3.0 [19], 2.95 [20], 1.56 [21] |

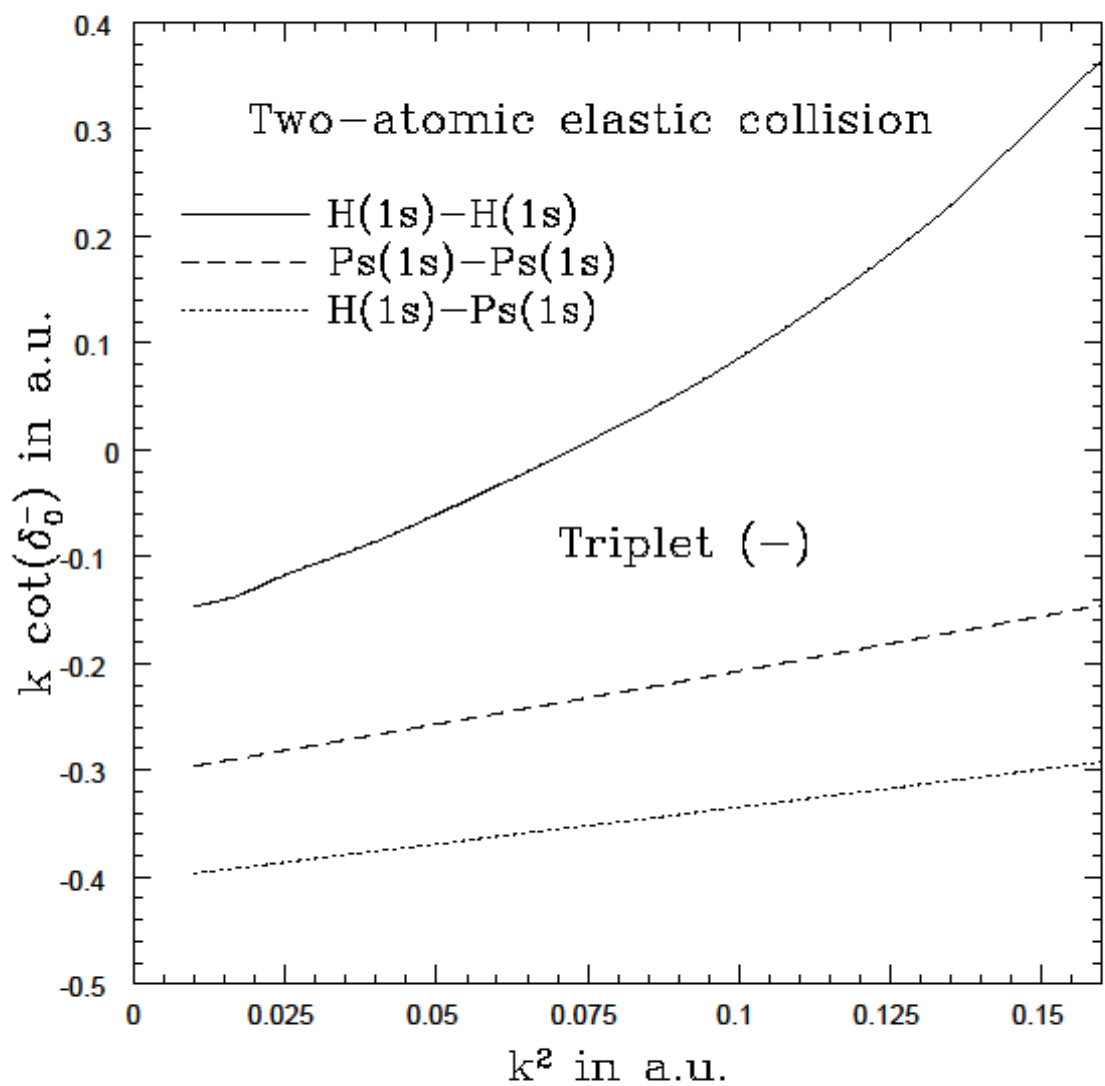

**Figure 3**. The plot of $k \cot \delta_0$ against $k^2$ in a.u. for the triplet (-) s-save elastic scattering in H(1s)-H(1s), H(1s)-Ps(1s) and Ps(1s)-Ps(1s) system using the new code.

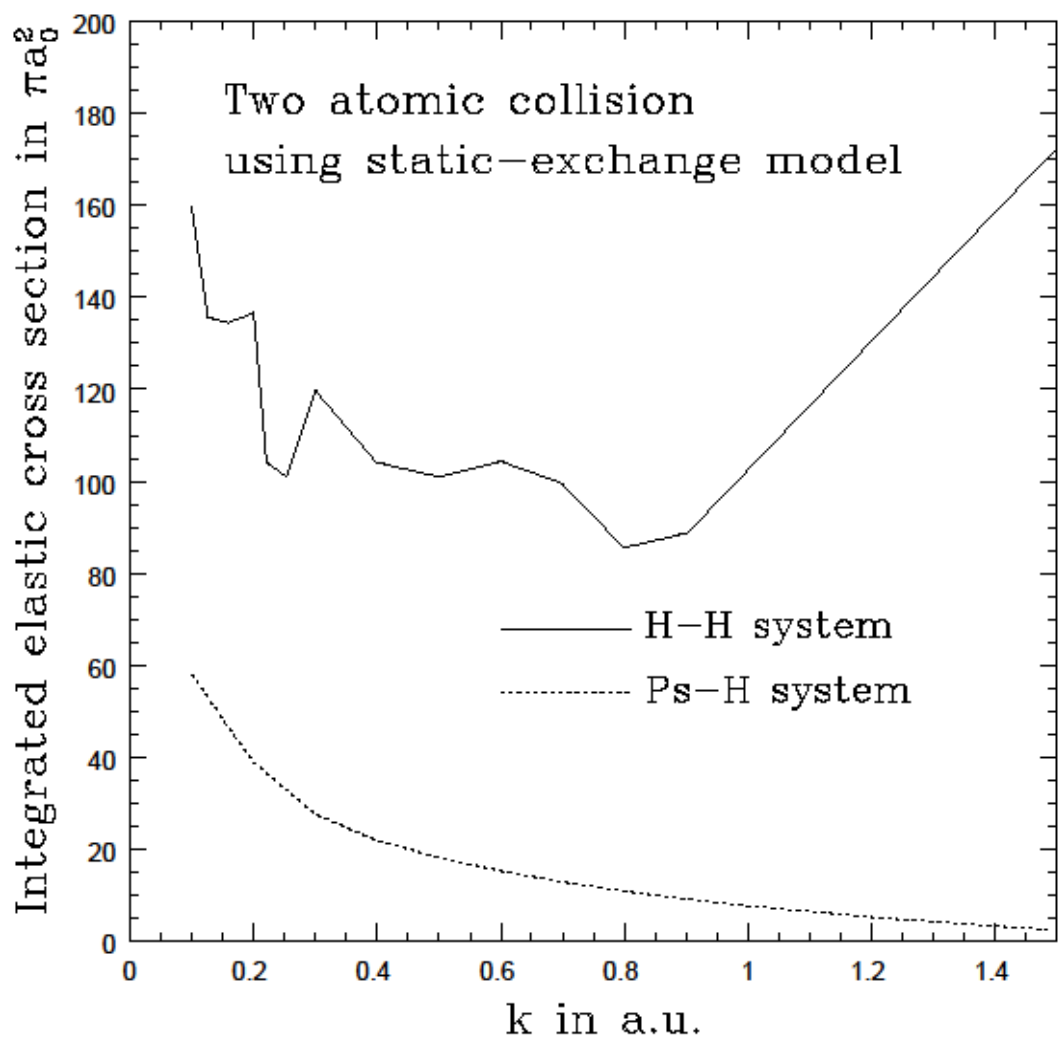

**Figure 4.** The variation of integrated elastic cross sections in $\pi a_0^2$ of H-H scattering.with the values of incident momentum k in a.u.